\documentclass[useAMS,usenatbib]{mn2e}

\usepackage{graphicx}
\usepackage{epsfig}
\usepackage{textcomp}

\renewcommand{\cos}{\mathrm{cos}}

\title[Wave propagation in the pulsar magnetosphere]
{On the radio wave propagation in the pulsar magnetosphere}
\author[V. S. Beskin and A. A. Philippov]{V. S. Beskin$^{1}$\thanks{E-mail:
beskin@lpi.ru} and A. A. Philippov$^{2}$
\\
$^{1}$P.N.Lebedev Physical Institute, Leninsky prosp., 53, Moscow, 119991, Russia\\
$^{2}$Moscow Institute of Physics and Technology, Dolgoprudny,
Moscow region, 141700, Russia}

\begin{document}

\date{Accepted, Received}


\maketitle

\label{firstpage}

\begin{abstract}
The key properties of the wave propagation theory in the magntosphere of radio pulsars based 
on the Kravtsov-Orlov equation are presented. It is shown that for radio pulsars with known 
circular polarization and the swing of the linear polarization position angle one can determine 
which mode, ordinary or extraordinary one, forms mainly the mean profile of the radio emission. 
The comparison of the observational data with the theory predictions demonstrates their good 
agreement.
\end{abstract}


\section{Introduction}

There are three main propagation effects in the magnetosphere of radio pulsars. 
They are refraction, cyclotron absorption, and the limiting polarization effects. 
The limiting polarization effect is related to the escape of radio emission from the region 
of dense plasma, where the propagation is well described within the geometrical optics 
approximation (in this case, the polarization ellipse is defined by the orientation of 
the external magnetic field in the picture plane), into the region of rarefied plasma, 
where the polarization of the wave remains almost constant along the ray. This process 
has been well-studied (Zheleznyakov, 1996; Kravtsov \& Orlov, 1980) and applied successfully, 
e.g., to the problem of solar radio emission (Zheleznyakov, 1970). 

However, in the theory of the pulsar radio emission such a problem has not been 
finally solved. Above the papers, where the position of the transition level 
$r_{\rm esc}$ between the domains of the geometrical optics 
and vacuum approximations was estimated (see, e.g., Cheng \& Ruderman 1979; Barnard, 1986), 
one can actually note only the papers by Petrova \& Lyubarskii (2000a) (where the 
problem in the infinite magnetic field was considered), and by Wang, Lai \& Han (2010)
(see Petrova, 2003, 2006 as well). But in all these papers the equations describing 
the evolution of the electric field of the waves were analysed. This approach cannot make 
the direct predictions concerning the polarization properties of the outgoing radiation.

We use another approach describing the propagation of electromagnetic waves in weakly 
inhomogeneous media, i.e., the method of the Kravtsov-Orlov equation that is well known 
in plasma physics and crystal-optics. It gives us the opportunity to write directly the 
equations on observable quantities, i.e., the position angle and the degree of circular 
polarization (CP). Also, the cyclotron absorbtion that naturaly occurs in finite magnetic field, 
and the possible linear transformation of waves can be easily included into consideration. 
 
The preliminary results based on the simple model of the magnetic field structure and energy 
distribution of particles flowing in the magnetosphere were already published by Andrianov \& 
Beskin (2010). It was shown that for radio pulsars with known circular polarization and the 
swing of the linear polarization position angle one can determine which mode, ordinary or 
extraordinary one, forms mainly the mean profile of the radio emission. Later, the arbitrary 
non-dipole magnetic field configuration, the drift motion of plasma particles, and their 
realistic energy distribution function were taken into account as well. The detailed
quantitative analysis of these effects will be presented in our separated paper. The goal 
of this Letter is in qualitative comparison of the main predictions of the theory with 
observational data. In our opinion, they are in a very good agreement.

\section{Theoretical predictions}\label{aba:sec1}

\subsection{On the number of outgoing waves}

Starting from the pioneer paper by Barnard and Arons (1986),
people discussed three waves propagating outward in the
pulsar magnetosphere. As shown in Fig. 1, for small enough
angles $\theta$ between the wave vector ${\bf k}$ and the
external magnetic field ${\bf B}$ two of them, $n_1$ and $n_2$,
are transverse waves, and the third mode $n_3$ corresponds
to plasma wave. The point is that in the most of papers
(see e.g. Melrose \& Gedalin, 1999) the waves properties were 
considered in the comoving reference frame in which the plasma 
waves propagating outward and backward are identical. But in 
the laboratory reference frame (in which the plasma moves with 
the velocity $v \approx c$) the latter wave is to propagate 
outward as well. Thus, in reality we have four waves propagating 
outward.

Moreover, as was demonstrated by Beskin, Gurevich \& Istomin (1993), 
it is the fourth wave that is to be considered as the O-mode 
in the pulsar magnetosphere. Indeed, as shown in Fig. 1,  for dense 
enough plasma in the radio generation domain for which $A_{\rm p} \gg 1$ 
where 
\begin{equation}
A_{\rm p}=\frac{\omega^2_{\rm p}}{\omega^2}<\gamma>,
\end{equation}
it is this wave that propagates as transverse one at large angles $\theta$ between 
${\bf k}$ and ${\bf B}$, i.e., at large distances from the neutron star. Here 
$\omega_{\rm p} = (4 \pi e^2 n_{\rm e}/m_{\rm e})^{1/2}$ is the plasma frequency, 
$n_{\rm e}$ is the concentration of particles, and $<\gamma>$ is the mean 
Lorentz-factor of the outflowing plasma. Two waves, for which the refractive 
index $n>1$, cannot escape from the magnetosphere as at large distances they 
propagate along the magnetic field lines (and due to Landau damping, see 
Barnard \& Arons, 1986). 

In the hydrodynamical limit one can easily obtain the dispersion curves shown in Fig. 1 
from the well-known dispersion equation for the infinite magnetic field (see, e.g., 
Petrova \& Lyubarskii, 2000) 
\begin{equation}
\left(1 - n^2 \cos^2\theta\right)
\left[1 - \frac{\omega_{\rm p}^2}{\omega^2 \gamma^3(1 - n v \cos\theta/c)^2}\right]
- n^2 \sin^2\theta = 0.
\end{equation}
For $\theta \ll \theta_*$ and for $\theta \gg \theta_*$, where
\begin{equation}
\theta_{*}=<\frac{\omega^2_{\rm p}}{\omega^2\gamma^3}>^{1/4},
\end{equation}
there are two transverse and two plasma waves, but for $A_{\rm p} \gg 1$
the nontrivial transformation from longitudinal to transverse wave takes place.
It means that for $A_{\rm p} \gg 1$ the mode $n_4$ can be emitted as 
a plasma wave, but it will escape from the magnetosphere as a transverse
one.

\begin{figure}
\centering\epsfig{figure=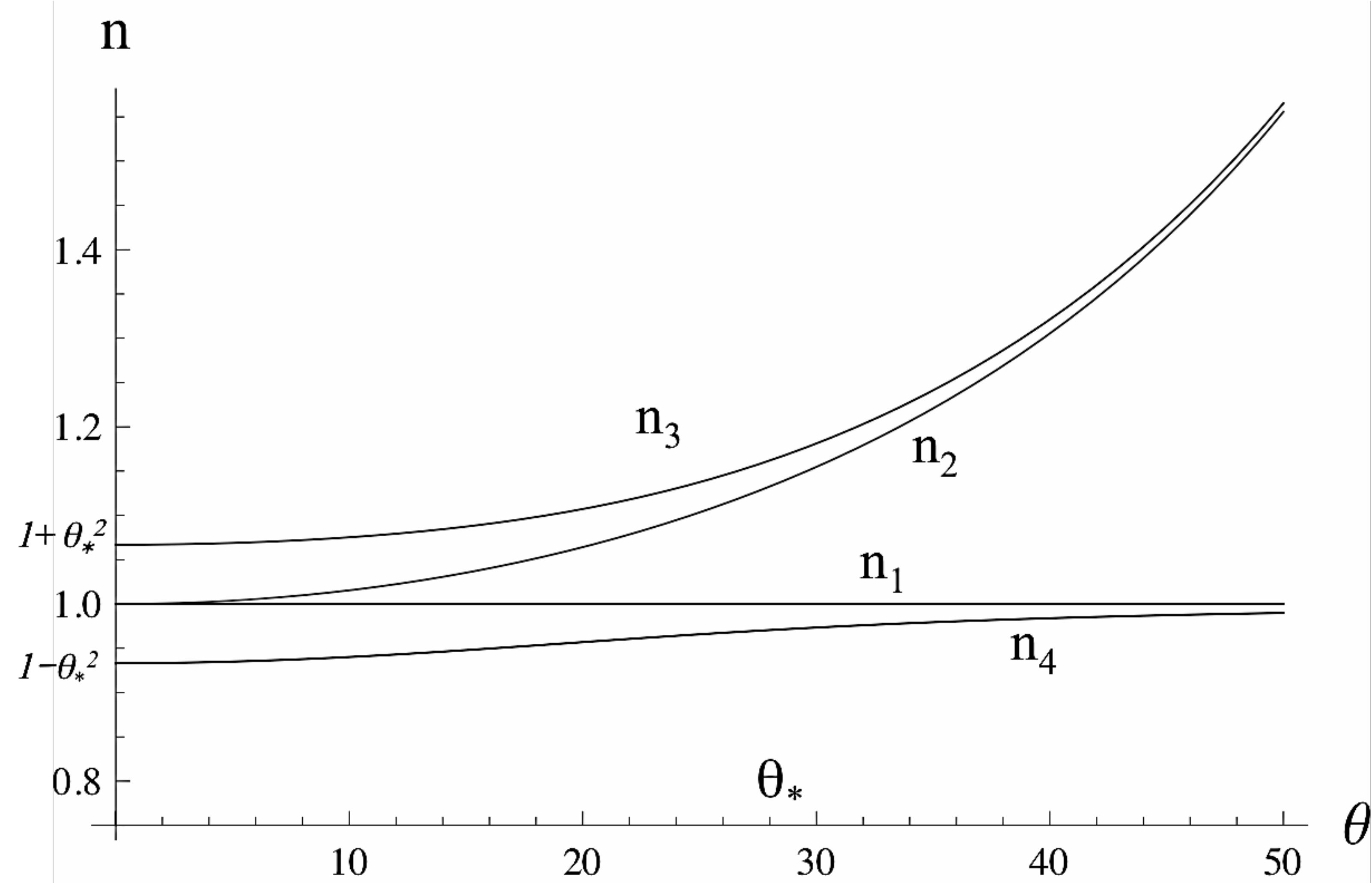,width=0.9\linewidth}
\caption{Dependence of the refractive indexes $n$ on angle $\theta$
for $A_{\rm p} \gg 1$. The lower branch corresponds to the O-mode} 
\label{fig1}
\end{figure}

\subsection{Kravtsov-Orlov equation}

The Kravtsov-Orlov equation describes the evolution of the complex angle 
$\Theta = \Theta_1 + i \Theta_2$,  where $\Theta_1$ is a position angle 
of the polarization ellipse, and $\Theta_2$ determines the circular polarization by the relation 
$V = I \, {\rm tanh}2\Theta_2$. Here $I$ is the intensity of the 
wave, and $V$ is the Stokes parameter. As was shown (Andrianov \& Beskin, 
2010, Beskin \& Philippov, 2011), the Kravtsov-Orlov equation can be 
rewritten as 
\begin{eqnarray}
\frac{{\rm d}\Theta_1}{{\rm d}l} = &&
\frac{\omega}{2c}{\rm Im}[\varepsilon_{x'y'}]
\nonumber \\
&& -\frac{1}{2}\frac{\omega}{c} \Lambda\cos[2\Theta_1-2\beta(l)-2\delta(l)]{\rm sinh}2\Theta_2,
\label{t1}\\
\frac{{\rm d}\Theta_2}{{\rm d}l} = &&
\frac{1}{2}\frac{\omega}{c}\Lambda\sin[2\Theta_1 - 2\beta(l)-2\delta(l)] {\rm cosh}2\Theta_2,
\label{t2}
\end{eqnarray}
where
\begin{equation}
\Lambda=\mp\sqrt{({\rm Re}[\varepsilon_{x'y'}])^2+\left(\frac{\varepsilon_{x'x'}-\varepsilon_{y'y'}}{2}\right)^2},
\end{equation}
where the signs corresponds to the regions before/after the cyclotron resonance and
\begin{equation}
\tan(2\delta)= - \frac{2{\rm Re}[\varepsilon_{x'y'}]}{\varepsilon_{y'y'}-\varepsilon_{x'x'}}.
\end{equation}
We would like to note, that in this equations the circular polarization is defined as it is common in radioastronomy (positive V corresponds to the co - clockwise rotating electric field vector for the observer).
Here $l$ is a coordinate along the ray propagation, and the angle $\beta(l)$ defines the 
orientation of the external magnetic field in the picture plane. In the geometric optics
region ($r < r_{\rm esc}$) the O-mode corresponds to polarization 
$\Theta_1 \approx  \beta + \delta$,
and X-mode corresponds to  $\Theta_1 \approx \beta + \delta + \pi/2$. 

Finally, $\varepsilon_{i'j'}$ are the components of plasma dielectric tensor in the reference 
frame where $z$-axis directs along the wave propagation and the external magnetic field lies 
in $xz$-plane. In comparison with model considered by Andrianov \& Beskin (2010), equations 
(\ref{t1})--(\ref{t2}) include into consideration nonzero components ${\rm Re}[\varepsilon_{x'y'}]$.
It allows us to take into account the electric drift of plasma particles. This effect just
corresponds to aberration considered by Blaskiewicz, Cordes \& Wasserman (1991). This effect 
was also considered by Petrova \& Lyubarskii (2000a), but for the infinite magnetic field only. 

Thus, knowing the magnetic field structure and plasma properties of the outgoing plasma
(i.e., knowing the dielectric tensor $\varepsilon_{i'j'}$) one can determine the observable 
physical quantities, namely, the Stokes parameter $V = I {\rm tanh}2\Theta_2(\infty)$ defining 
the CP and the position angle $p.a. = \Theta_1(\infty)$ characterizing the orientation of the 
polarization ellipse. This approach is valid in the quasi-isotropic case when there are two 
small parameters, i.e., the general WKB condition $1/kL \ll 1$ and the condition
$\Delta n/n_{1,4} \ll 1$, where $\Delta n = n_{1}-n_{4}$. 

\subsection{Main predictions of the propagation theory}

As the refractive index $n_4$ differs from unity, the appropriate ordinary mode is to
deflect from the magnetic axis until $\theta  \leq \theta_{*}$. As was already mentioned, 
for the O-mode this effect takes place until $A_{\rm p} > 1$, i.e., for small enough 
distances from the neutron star $r < r_{\rm A}$, where
\begin{eqnarray}
r_{\rm A} \approx  10^{2} R \,  
\lambda_{4}^{1/3}  \,
\gamma_{100}^{1/3} \,
B_{12}^{1/3}  \,
\nu_{\rm GHz}^{-2/3} \,
P^{-1/3}.
\label{rA}
\end{eqnarray}
Here $R$, $P$, and $B_{12}$ are the netron star radius, rotation period (in s), and magnetic
field (in $10^{12}$ G), respectively. Accordingly,  $\gamma_{100} = \gamma/100$, $\nu_{\rm GHz}$ 
is the wave frequency in GHz, and 
$\lambda_{4} = \lambda/10^{4}$, where $\lambda = n_{\rm e}/n_{\rm GJ}$ is the multiplicity 
of the  particle creation near magnetic poles ($n_{\rm GJ} = \Omega B/2 \pi c e$ is the 
Goldreich-Julian concentration). On the other hand, the transverse extraordinary 
wave with the refractive index \mbox{$n=1$} (X-mode) is to propagate freely. As the radius 
$r_{\rm A}$ is much smaller than the escape radius $r_{\rm esc}$ (Cheng \& Ruderman, 
1979, Andrianov \& Beskin, 2010)
\begin{eqnarray}
r_{\rm esc} \approx  10^{3} R \,  
\lambda_{4}^{2/5} \,
\gamma_{100}^{-6/5} \,
B_{12}^{2/5} \,
\nu_{\rm GHz}^{-2/5} \,
P^{-1/5},
\label{resc}
\end{eqnarray}
one can consider the effects of refraction and limiting polarization separately.
In particular, this implies that one can consider the propagation of waves in the
region $r \sim r_{\rm esc}$ as rectilinear.
 
\begin{figure}
\centering\epsfig{figure=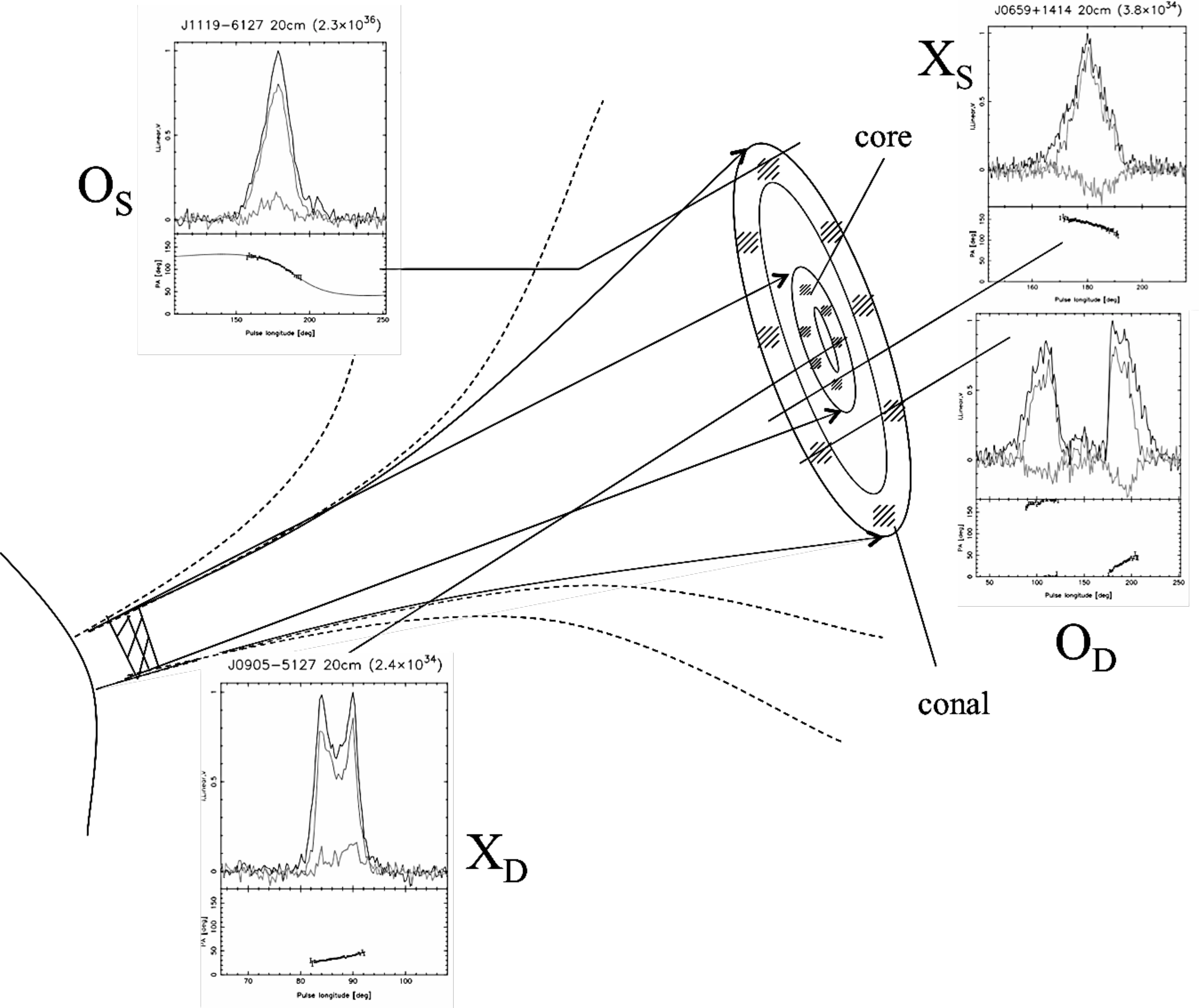,width=0.95\linewidth}
\caption{Propagation of X- and O-modes in the pulsar magnetosphere. The lower curve on each top panel of each picture 
indicates V, on the bottom panel p.a. curve is shown.} 
\label{fig2}
\end{figure}

As a result, as shown in Fig. 2, if two modes are generated at the same distance 
$r_{\rm gen} < r_{\rm A}$ from the neutron star where $A_{\rm p}(r_{\rm gen}) > 1$, 
the directivity pattern of the O-mode is to be wider than that for the X-mode.
Hence, it is logical to assosiate the X-mode with the core component of the 
directivity pattern. On the other hand, the O-mode is to form the conal component.  
E.g., assuming the constant plasma density across the flow (more realistic regime was 
considered by Petrova \& Lyubarskii, 2000),  Beskin, Gurevich \& Istomin (1993)
obtained for the radio window width $W_{\rm r}$
\begin{eqnarray}
W_{\rm r}^{({\rm X})} & \approx & 
3.6^{\circ} \, P^{-3/4} \, \nu_{\rm GHz}^{-1/2}, 
\label{W1} \\
W_{\rm r}^{({\rm O})} & \approx & 
7.8^{\circ} \, P^{-0.43} \, \nu_{\rm GHz}^{-0.14},
\label{W2} \\
W_{\rm r}^{({\rm O})} & \approx & 
10^{\circ} \, P^{-0.5} \, \nu_{\rm GHz}^{-0.29}.
\label{W3}
\end{eqnarray}
For the O-mode, the theory gives two radiation beam widths, depending on the distribution 
of the energy release power in height. 

The main new theoretical result based on the analysis of Kravtsov-Orlov equations 
(\ref{t1})--(\ref{t2}) is that for large enough derivative ${\rm d}(\beta+\delta)/{\rm d} l$, the sign of the 
Stokes parameter $V$ is to be determined not by the poorly defined nondiagonal 
components of the dielectric tensor $\varepsilon_{x'y'}$, when $V_0/I = \pm 1/q$ 
(Zheleznyakov, 1977) and
\begin{equation}
q = i \, \frac{\varepsilon_{y'y'} - \varepsilon_{x'x'}}{2\varepsilon_{x'y'}},
\end{equation}
but by the sign of the derivative ${\rm d}(\beta+\delta)/{\rm d} l$. It takes place when one 
can neglect the first term in the r.h.s. of Eqn. (\ref{t1}) in comparison with the 
derivative ${\rm d}\Theta_1/{\rm d} l  \approx {\rm d}(\beta+\delta)/{\rm d} l$. E.g., for 
$r \ll r_{\rm esc}$ ($|V| \ll 1$):
\begin{equation}
\frac{V}{I} \approx 
\frac{1}{|q|} \cdot \frac{{\rm d}(\beta+\delta)/{\rm d}x}{|v_{\parallel}/c - \cos\theta|}\, 
\frac{1}{\cos[2\Theta_1-2\beta(l)-2\delta(l)]}.
\label{main}
\end{equation}
Here $x = \Omega l/c$ is dimensionles distance along the ray.

Hence, for large enough ${\rm d}(\beta+\delta)/{\rm d}x \sim 1$ 
(the total turn $\Delta (\beta+\delta) \sim 1$ within the light cylinder $R_{\rm L} = c/\Omega$), 
and for small angle of propagation $\theta \ll 1$ through the relativistic plasma 
($v_{\parallel}/c \sim 1$) the Stokes parameter $V$ (\ref{main}) is indeed much larger 
than $V_0/I = \pm 1/q$ resulting from standard evaluation. Moreover, numerical calculations (Beskin \& Philippov, 2011) 
shows that the sign of the derivative 
${\rm d}(\beta+\delta)/{\rm d} l$ is opposite to the sign of the derivative ${\rm d}p.a./{\rm d} \phi$. 

As a result, one can formulate the following predictions:
\begin{itemize}
\item
For the X-mode ($\Theta_1 \approx \beta + \delta + \pi/2$) the theory predicts the SAME signs 
of the circular polarization $V$ and the derivative ${\rm d} p.a./{\rm d}\phi$.
\item
For the O-mode ($\Theta_1 \approx \beta + \delta$) the signs of the circular 
polarization $V$ and the derivative ${\rm d} p.a./{\rm d}\phi$ are to be OPPOSITE.
\item 
For radio pulsar with the tripple mean pulses we have to see the O-mode first, than X- 
and again the O-mode because, as was shown in Fig. 2, the O-mode deviates from magnetic 
axis. 
\item 
In general, the trailing part of the main pulse can be absorbed (see Dyks, 
Wright \& Demorest, 2010 as well).
\item 
The leading component can be absorbed only if the polarization formes near the light
cylinder ($r_{\rm esc} \sim R_{\rm L}$). In this case the $p.a.$ is to be 
approximately constant in given mode as the magnetic field here is approx. homogeneous
(A.Spitkovsky, private communication).
\item 
Statistically, we expect single profiles for the X-mode (if only the X-mode is observed) 
and the double ones for the O-mode.
\end{itemize}
In the following section we will try to demonstrate that these predictions are indeed in good
agreement with observational data.  
\section{Comparison with observations}
\subsection{Statistics on single-mode pulsars}

In Table~\ref{table1} we collected more than 70 pulsars from two reviews by
Weltevrede \& Johnston (2008) and Hankins \& Rankin (2010) for
which both the values of $p.a.$ and $V$ were well-determined. As was predicted, 
most pulsars with double (D) mean profile corresponds to the O-mode (the opposite signs 
of  $V$ and  ${\rm d} p.a./{\rm d}\phi$), and the most pulsars with single (S) mean profiles -- 
to the X-mode (the same signs of  $V$ and  ${\rm d} p.a./{\rm d}\phi$).  Moreover, statistically 
the O-mode pulsars have wider mean pulses than X-mode, their values being in good agreement
with theoretical predictions (\ref{W1})--(\ref{W3}).

\begin{table}
\caption{Statistics of pulsars with known circular polarization $V$ and position angle swing.
Pulsar period $P$ is in seconds, and the window width $W_{50}$ is in degrees.}  
\centering
 \begin{tabular}{|c|c|c|c|c|}
 \hline
${\rm Profile}$& ${\rm O_{S}}$ & ${\rm O_{D}} $ & ${\rm X_{S}} $ & ${\rm X_{D}} $\\
\hline
Number&6&23&45&6\\
\hline
$\sqrt{P}W_{50}$&6.8$\pm$ 3.1&10.7$\pm$ 4.5&6.5$\pm$ 2.9&5.3$\pm$ 3.0\\
\hline
\end{tabular}
\label{table1} 
\end{table}
\subsection{Pulsars with almost constant $p.a.$}

In Table~\ref{table2} we collected ten pulsars with almost constant $p.a.$ from the papers by 
Johnston et al. (2007) ([1] in the reference column),  Mitra and Rankin (2010) ([2]), and 
Johnston et al. (2008) ([3]). One can note that their rotational periods are smaller than 
1 s, i.e., the typical pulsar period. Also four of them (indicated by plus in the comment column) 
show the "flatting" of the $p.a.$ swing with the frequency decreasing. The interpretation of 
these effects can be easily given basing on the estimate of the escape radius $r_{\rm esc}$
(\ref{resc}), where the  polarization of the radio emission forms (Andrianov \& Beskin, 2010).
As $r_{\rm esc}/R_{\rm L} \propto P^{-6/5} \nu^{-2/5}$, one can see that for small pulsar periods and lower 
frequencies the outgoing polarization forms closer to the light cylinder where magnetic 
field of neutron star, as was already stressed, is almost homogeneous. In this case the $p.a$ 
of the outgoing radiation is to be approximately constant within the mean pulse. 
This effect was firstly introduced by Barnard (1986), the main distinction of our theory is in self-consistent 
definition of $r_{\rm esc}$ basing on the solution of equations (\ref{t1})--(\ref{t2}).

\begin{table}
\caption{Pulsars with almost constant p.a. Here P is in seconds, $\dot{P}$ is in $10^{-15}$, 
and magnetic field $B$ on the surface of neutron star is in $10^{12}$ G}  
\centering
 \begin{tabular}{|c|c|c|c|c|c|}
  \hline
${\rm PSR}$& $P(s)$ & $\dot{P}  $ & $B_{12}$ & comment & reference\\
\hline
J0543$+$2329&   0.25 &    15.4            &   2.0           &   +     & [3] \\
\hline
J0738$-$4042&   0.37 &     1.6            &   0.8           &   +     & [1] \\
\hline
J0837$-$4135&   0.75 &     3.5            &   1.6           &         & [1] \\
\hline
J1559$-$4438&   0.26 &     1.0            &   0.5           &   +     & [3] \\
\hline
J1735$-$0724&   0.42 &     1.2            &   0.7           &         & [1] \\
\hline
B1907$-$03  &     0.50 &   2.2            &   1.0           &         & [2] \\
\hline
J1915$+$1009&   0.40 &     1.5            &   2.5           &         & [1] \\
\hline
J1937$+$2544&   0.20 &     0.6            &   0.4           &   +     & [1] \\
\hline
\end{tabular}
\label{table2} 
\end{table}

\subsection{Pulsars with tripple mean profiles}

Below we analyse the profiles of several tripple pulsars presented in review by Johnston et al. (2007)
for which both $p.a.$ and $V$ are well-determined at any way at one frequency, 693, 1374, or 3100 MHz.

For pulsar PSR J0452$-$1759 the triple structure is seen at 693 MHz only. For this reason, it is
not surprizing that the $90^{\circ}$-switch of the $p.a.$ from O-mode to X-mode and return is seen 
on this frequency only. The CP is well-determined in the trailing subpulse only, where its sign, as was 
predicted, corresponds to O-mode.

The tripple structure of the pulsar PSR J0738$-$4042 is seen on the $p.a.$ profile at 691 MHz only.
Nevertheless, as the $p.a.$ is approximately constant at low frequency (see Table 2), one can assume 
that the leading part of the profile is absorbed. Then, the leading subpulse is to correspond the X-mode, 
and the trailing one -- to the O-mode (the appropriate negative CP is seen in this subpulse only). 

In pulsar PSR J1559$-$4438 the tripple structure O-X-O is well-seen at all frequencies 691, 1374, and
3100 MHz, but in the mean profile the conal component is detected at the frequency 3100 MHz only. The negative
CP for core subpule corresponding to the X-mode is seen at all frequencies. 

The CP of the tripple pulsar PSR J2048$-$1616, as was predicted, corresponds to O-X-O structure.
On the other hand, the switch of the $p.a.$ is absent, that can be caused, in our opinion, by the averaging technique 
and should be checked in the individual profiles as well (one can see the influence of the averaging technique 
on the CP profile in Karastergiou et al., 2003).


Thus, as we see, the properties of the pulsars with triple mean profiles are in good agreement with the theoretical
predictions. The central (core) components of the mean profiles, in general, connect with the X-mode propagating freely, 
while the conal parts correspond to the O-mode deflecting via the refraction from the magnetic axis. 

\subsection{Pulsars with interpulses}

Here the same analysis for pulsars with interpulses from paper by Keith et al. (2010) is given. The observations 
were made at the frequency 1.4 GHz.

The main pulse of pulsar PSR J0627$+$0706 is to be connected with the O-mode (it has the opposite signs of $p.a.$ and
$V$). The trailing subpulse of the double profile, as is well seen from the $p.a.$ swing (the observable subpulse 
corresponds to the first half of the S-shape curve) is absorbed. In the interpulse the circular polarization is not 
high enough to determine modes.

The central part of the main pulse of pulsar PSR J1549$-$4848 is formed by the X-mode (the same signs of the $p.a.$ and $V$). 
In the leading and the trailing parts of the main pulse the CP is too low to say anything about the conal component.
The interpulse has the triple structiure, the sequence of the modes, as was predicted, being O-X-O. This is clear
not only from the negative-positive-negative sequence of the Stokes parameter $V$ but from the singularity of the
$p.a.$ at the phase $\phi=-85^{\circ}$. 

The main pulse of pulsar PSR J1722$-$371 is formed mainly by the X-mode. 
The CP of the interpulse is too low to determine the modes.

The double main pulse of the pulsar PSR J1739$-$2903 is to be interpreted as the sequence of the O- and X-modes.
It is clear both from the negative-to-positive change of the Stokes parameter $V$ and from the $90^{\circ}$ jump in $p.a.$
The trailing (O-mode) component is absorbed because, as for PSR J0627$+$0706, the leading O-mode subpulse corresponds 
only to the first half of the S-shape curve of the $p.a.$.
The interpulse is formed by the X-mode.

In the main pulse of pulsar PSR J1828$-$1101 the value $V$ is not high enough to determine the mode.
The interpulse is formed mainly by the O-mode, the trailing subpulse being absorbed (here again the
$p.a.$ of the visible pulse corresponds to the first half of the S-shape curve).

Thus, for radio pulsars with interpulse the interpretation on the ground of the theory under
consideration is reasonable as well.

\subsection{Milliseconds pulsars}

Finally, we consider the millisecond pulsars which detailed polarization characteristics
were presented recently by Yan et al., (2011). The observations were made at the frequency 
1369 MHz.

Pulsar PSR J0437$-$4715 has the multiple mean profile. Its leading and the trailing parts 
(i.e., the conal component) definitely corresponds to the O-mode. On the other hand, both
the $90^{\circ}$ jump in $p.a.$ and the change of the Stokes parameter sign in the core
component shows that the central subpulse consists not only of the X-mode, but of the O-mode 
as well. 
 
The leading and the trailing components of the double profile of pulsar PSR J1022$+$1001, in agreement with the 
prediction, are formed by the O-mode. The peak on the $p.a.$ curve in the center of the mean profile could be 
connected with the core X-mode, but its intensity is too low to be seen on the mean pfofile. 

Pulsar PSR J1045$-$4509 has the triple main profile. It has different signs of the circular polarization 
in the core and conal components. But the position angle swing is irregular, which prevent us to determine
the modes.

In pulsar PSR J1600$-$3053 the core component is definitely formed by the X-mode (the same signs of the Stokes
parameter $V$ and the derivative ${\rm d} p.a./{\rm d}\phi$). Then, the conal component is to be connected with
the O-mode as its $p.a.$ curve locates approx. $90^{\circ}$ higher that that for the core component. The 
circular polarization is too low to confirm this point. 

The circular polarization of the central part of the double profile of pulsar PSR J1603$-$7202 is different from the
conal ones. The irregular character of the $p.a.$ swing does not allow us to determine the mode. 

Pulsar PSR J1643$-$1224 has the single main profile which can be easily interpreted as standard 
O-X-O sequence with the absorbed trailing part. Indeed, the polarization of the leading part definitely corresponds 
to the O-mode (the different signs of $V$ and ${\rm d} p.a./{\rm d}\phi$), while the trailing part is to be
connected with the X-mode (the same signs of $V$ and ${\rm d} p.a./{\rm d}\phi$). Moreover, the derivative
${\rm d} p.a./{\rm d}\phi$ is much larger in the trailing part. This implies that the visible pulse corresponds
the first half of the S-shape curve.

Pulsar PSR J1713$+$0747 has multicomponent profile containing both modes. It has different signs of the circular 
polarization for different branches of the p.a. curves.

Pulsar PSR J1732$-$5049 has actually zero circular polarization which does not allow us to determine the modes. 
Nevertheless, one can see that the $p.a.$ of the core and conal parts belongs to the different modes.

The profile of pulsar PSR J1909$-$3744 is quite similar to PSR J1643$-$1224. Hence, it can be easily interpreted 
as O-X-O sequence with the absorbed trailing part as well.

Radiation of pulsar PSR J2129$-$5721 is to be connected with the core conponent formed by the X-mode (the same signs 
of $V$ and ${\rm d} p.a./{\rm d}\phi$). Two subpulses of the mean profile can be easily explained by the central passage 
through the directivety pattern (see Fig. 2).

In addition, the double profile pulsars PSR J0613$-$0200, J0711$-$6830 have the same signs of the Stokes parameter $V$ 
in both subpulses. But it is impossibe to connect them  with the O-mode because of the irregular swing ofthe $p.a.$.
Finally, pulsars J1024$-$0719,  J1730$-$2304, J1744$-$1134, J1824$-$2452, J1857$+$0943, 2124$-$3358, and J2145$-$0750 
have irregular structure or actually zero circular polarization which does not allow us to analyse their properties.

Thus, in that cases when the $p.a.$ swing and the Stokes parameter $V$ are regular enough,
the main properties of the mean profiles can be easily interpreted within the theory
under consideration.

\section{Acknowledgments}

We thank A.V.~Gurevich and Y.N.~Istomin for their interest and support, J.~Dyks,
A. Jessner, D.~Mitra, M.V.~Popov, B.~Rudak and H.-G.~Wang for useful discussions, 
and an anonymous referee for valuable suggestions 
that greatly improved the quality of this article. This 
work was partially supported by Russian Foundation for Basic Research (Grant no. 
11-02-01021).


\begin{thebibliography}{}
\bibitem[\protect\citeauthoryear{Andrianov \& Beskin}{2010}]{Beskin-10}
Andrianov~A.S., Beskin~V.S., 2010, Astron. Letters, 36, 248

\bibitem[\protect\citeauthoryear{Barnard}{1986}]{Barnard}
Barnard~J.J., 1986, ApJ, 303, 280

\bibitem[\protect\citeauthoryear{Barnard \& Arons}{1986}]{BarnardArons}
Barnard~J.J., Arons~J., 1986, ApJ, 302, 138

\bibitem[\protect\citeauthoryear{Beskin \& Philippov}{2011}]{bph} 
Beskin~V.S., Philippov~A.A., 2011 (in preparation)

\bibitem[\protect\citeauthoryear{Beskin \& Gurevich \& Istomin}{1993}]{bgi2} 
Beskin~V.S., Gurevich~A.V., Istomin~Y.N., 1993, Physics of the Pulsar 
Magnetosphere, Cambridge University Press, Cambridge

\bibitem[\protect\citeauthoryear{Blaskiewicz, Cordes \& Wasserman}{1991}]{bcw} 
Blaskiewicz~M., Cordes J.M., \& Wasserman I., 1991, ApJ, 370, 643

\bibitem[\protect\citeauthoryear{Cheng \& Ruderman}{1979}]{Cheng-79}
Cheng~A.F., Ruderman~M.A., 1979, ApJ, 229, 348

\bibitem[\protect\citeauthoryear{Dyks, Wright \& Demorest}{2010}]{Dyks-10}
Dyks~J., Wright~G.A.E., Demorest~P.B., 2010, MNRAS, 405, 509

\bibitem[\protect\citeauthoryear{Hankins \& Rankin}{2010}]{Rankin-10}
Hankins~T.H., Rankin~J.M., 2010, AJ, 139, 168

\bibitem[\protect\citeauthoryear{Johnston \& Kramer}{2005}]{Johnston-05}
Johnston~S., Karastergiou A., Mitra D., Gupta Y., 2008, MNRAS, 388, 261

\bibitem[\protect\citeauthoryear{Johnston \& Kramer}{2007}]{Kramer-07}
Johnston~S., Kramer~M., Karastergiou~A., Hobbs~G., Ord~S., Wallman~J., 2007, MNRAS, 381, 1625

\bibitem[\protect\citeauthoryear{Karast}{2003}]{Karastergiou}
Karastergiou~A., Johnston~S., Kramer~M., 2003, A\&A, 404, 325

\bibitem[\protect\citeauthoryear{Johnston \& Kramer}{2010}]{Kramer-10}
Keith~M.J., Johnston~S., Weltrvrede~P., Kramer~M., 2010, MNRAS, 402, 745

\bibitem[\protect\citeauthoryear{Kravtsov \& Orlov}{1990}]{ko} 
Kravtsov~Yu.A., Orlov~Yu.I., 1990, Geometrical Optics of Inhomogeneous Media, Springer,
Berlin

\bibitem[\protect\citeauthoryear{Melrose \& Gedalin}{1999}]{Melrose-99}
Melrose~D.B., Gedalin~M.E., 1999, ApJ, 521, 351

\bibitem[\protect\citeauthoryear{Mitra \& Rankin}{2010}]{Mitra-10}
Mitra~D., Rankin~J.M., 2010, ApJ, 727, 92

\bibitem[\protect\citeauthoryear{Petrova}{2001}]{Perova03}
Petrova~S.A.,  2003, A\&A, 408, 1057

\bibitem[\protect\citeauthoryear{Petrova}{2006}]{Perova06}
Petrova~S.A.,  2006, MNRAS, 368, 1764

\bibitem[\protect\citeauthoryear{Petrova \& Lyubarskii}{2000a}]{Petrova-00}
Petrova~S.A., Lyubarskii~Yu.E., 2000, A\&A, 355, 1168

\bibitem[\protect\citeauthoryear{Wang \& Lai \&Han}{2010}]{Lai-10}
Wang~C., Lai~D., Han~J., 2010, MNRAS, 403, 2

\bibitem[\protect\citeauthoryear{Weltvrede \& Johnston}{2008}]{Welt-79}
Weltrvrede~P., Johnston~S., 2008, MNRAS, 391, 1210

\bibitem[\protect\citeauthoryear{Yan et al}{2011}]{Yan}
Yan~W.M., Manchester~R.N., van Straten~W. et al., 2011, arXiv:1102.2274

\bibitem[\protect\citeauthoryear{Zheleznyakov}{1964}]{Zhel-64}
Zheleznyakov~V.V., 1970, Radio Emission of the Sun and Planets, Pergamon, Oxford

\bibitem[\protect\citeauthoryear{Zheleznyakov}{1977}]{Zhel-77}
Zheleznyakov~V.V., 1996, Radiation in Astrophysical Plasmas, Springer, Berlin
\end{thebibliography}
\end{document}